# Calculating the noise flux of spontaneous emission radiated from Hydrogen-like atoms by means of vibrational Hamiltonian and ignoring from ambient effects

## J. Jahanpanah

Physics Faculty, Kharazmi University, 49 Mofateh Ave, 15719-14911, Tehran, Iran

E-mail: jahanpanah@khu.ac.ir

**Abstract**

A theoretical model is introduced for constructing the vibrational Hamiltonian of Hydrogen-like atoms (HLA). The Hamiltonian is then used to derive the vibrational motion equations of HLA in Heisenberg picture. The Langevin equation will ultimately be formed after adding the dissipative term and fluctuating (Langevin) force according to the fluctuation-dissipation theorem (FDT). The positional noise flux is then defined as the correlation function of fluctuations that happens for the electron position during its rather fast vibrational oscillations ($\omega_{vib} \approx 10^{15}\ Hz$). On the other hand, the positional fluctuations led to the fluctuations in the potential and kinetic energies of oscillating electron so that the appearance of the potential and kinetic noise fluxes is vulnerable. The positional, potential, and kinetic noise fluxes of oscillating electron will be determined by solving the Langevin equation in frequency domain. It is finally demonstrated that the potential and kinetic noise fluxes commonly act as an internal source for producing the external noise flux emitted from HLA in the form of spontaneous emission with a Lorentzian profile. In contrast with all previous procedures, no ambient effect has been involved to describe the forming mechanism of spontaneous emission for the first time.

## 1. Introduction

The spontaneous emission (SPE) was well analyzed by Milonni as the simultaneous interaction of electron with the vacuum field and the field of radiation reaction [1]. Subsequently, many articles have been published to describe SPE by implementing the different external sources such as interacting with the zero-point energy, thermal photons, radiation reaction field, virtual photons, and so on [2-6]. We have here interpreted SPE from a different aspect by looking at the atom from inside and constructing its vibrational Hamiltonian away from the above exotic effects. Our model is based on the energy eigenvalue $E_n = (n+1/2)\hbar\omega_0$ of a simple harmonic oscillator (SHO), which immediately converted to its Hamiltonian operator $\hat{H}_0 = (\hat{N}+1/2)\hbar\omega_0$ by substituting the number operator $\hat{N}$ for the energy level number $n$. This simple case persuaded us to extend this method for the more complicated systems such as atoms and molecules.

Recently, the vibrational Hamiltonian of diatomic molecules has been formed up to the third power of $\hat{H}_0$ as $\hat{H}^{(3)} = \hat{H}_0 + \gamma_2 \hat{H}_0^2 + \gamma_3 \hat{H}_0^3$ [7] in order that the second and third-order quantization coefficients $\gamma_2$ and $\gamma_3$ were determined by inspiring from the stability theory of lasers [8]. The vibrational motion equations of diatomic molecules have then been derived by applying $\hat{H}^{(3)}$ into the Heisenberg equation. The solutions led to the useful information about the molecular vibration frequencies in the different energy levels of Morse potential, and the last vibrational stable level (dissociation level) [7].

The above method is here extended to cover the more complicated case of Hydrogen-like atoms (HLA) due to the Coulomb potential and its divergent behaviour (singularity) at the origin. It is then turned out that the problem is sorted out by excluding the zero-point energy ($E_0 = \hbar\omega_0/2$) from the energy spectrum of SHO. Similarly, by substituting the vibrational Hamiltonian of HLA into the Heisenberg equation, an equation is derived for the vibrational motion of electron around the rest nuclei which acts like a quantum harmonic oscillator with the oscillatory frequencies in consistent with Bohr's classical model.

On the other side, the sudden transition of oscillating electron from the upper to a lower atomic energy level is an internal vulnerable matter which led to the external radiation of SPE from HLA with a Lorentzian profile. As a result, the Langevin equation, which describes the motion of a damped oscillator (Langevin oscillator), is formed by adding the damping term and the fluctuation (Langevin) force into the vibrational equation of HLA according to FDT [9, 10]. The solution of Langevin equation gives the fluctuations that imposed to the both potential and kinetic energies of HLA during the fast vibrational oscillation of its electron. The correlation function of Langevin force is then used to calculate the noise flux of SPE which radiated from HLA during the atomic transition. At the end, it will be demonstrated that the noise flux of SPE is not only provided by the potential and kinetic noise fluxes of oscillating electron but also by their interchange noise fluxes according to the flux conservation Law.

## 2. Constructing the vibrational Hamiltonian of HLA

The first step initiates with the vibrational energy eigenvalue of HLA which was derived from solving the time-independent equation of Schrodinger associated with a two-particle system as $E_{vib} = E_n = -0.5\mu c^2 (Z\alpha)^2 / n^2$, in which $\mu$, $c$, $Z$, $\alpha$, and $n = 1, 2, 3, \ldots$ are the reduced mass, light speed, atomic number, fine constant, and energy level number, respectively [11]. The vibrational energy eigenvalue $E_{vib}$ is now expanded around the lowest value $n = 1$ as

$$E_{vib} = 0.5\mu c^2 (Z\alpha)^2 \sum_{k=0}^{\infty} (-1)^{k+1} (k+1)(n-1)^k \qquad (1)$$

On the other hand, there is no connection between $E_{vib}$ and the usual energy eigenvalue of SHO $E_n = (n+1/2)\hbar\omega_0$, except the non-integer value $1/2$ associated with the zero-point energy is excluded. Therefore, an infinite ladder energy spectrum with the equal steps $\hbar\omega_0$ remains for the energy spectrum of SHO as

$$E_n = (n-1)\hbar\omega_0, \qquad (2)$$

which is in consistent with the expanded eigenvalue $E_{vib}$ given by (1). It is emphasized that an infinite ladder with the equal steps $\hbar\omega_0$ can be demonstrated by $n\hbar\omega_0$ or equivalently by $(n-1)\hbar\omega_0$.

The next step is to substitute the number operator $\hat{N}$ for the energy level number $n$ in the both energy eigenvalues (1) and (2) to derive the corresponding Hamiltonian operators in the forms

$$H_{vib} = 0.5\mu c^2 (Z\alpha)^2 \sum_{k=0}^{\infty} (-1)^{k+1} (k+1)(\hat{N}-1)^k \qquad (3)$$

and

$$\hat{H}_0 = (\hat{N}-1)\hbar\omega_0. \qquad (4)$$

The last step is to form the vibrational Hamiltonian of HLA ($\hat{H}_{vib}$) in term of the Hamiltonian of SHO ($\hat{H}_0$) by substituting ($\hat{N}-1$) from (4) into (3) as

$$\hat{H}_{vib} = 0.5\mu c^2 (Z\alpha)^2 \sum_{k=0}^{\infty} (-1)^{k+1}(k+1)\left(\frac{\hat{H}_0}{\hbar\omega_0}\right)^k, \tag{5}$$

in which all the expansion coefficients will simultaneously be determined by calculating the value of fundamental frequency $\omega_0$ in the following section.

## 3. The vibrational Hamiltonian in the linear regime

The linear regime of HLA-Hamiltonian is corresponding to $k \leq 1$ in the general relation (5) as

$$\hat{H}_{vib}^{(1)} = 0.5\mu c^2 (Z\alpha)^2 \left(-1 + \frac{2\hat{H}_0}{\hbar\omega_0}\right). \tag{6}$$

The first-order vibrational equations of motion are now derived for the mean values of relative position operator $<\hat{x}>$ and linear momentum operator $<\hat{p}>$ by using the well-known Heisenberg equation in quantum mechanics as [12]

$$\frac{d<\hat{x}(t)>}{dt} = i\hbar^{-1} <[\hat{H}_{vib}^{(1)}, \hat{x}(t)]> = \frac{c^2(Z\alpha)^2}{\hbar\omega_0} <\hat{p}(t)> \tag{7}$$

and

$$\frac{d<\hat{p}(t)>}{dt} = i\hbar^{-1} <[\hat{H}_{vib}^{(1)}, \hat{p}(t)]> = \frac{-\mu^2 c^2 (Z\alpha)^2 \omega_0}{\hbar} <\hat{x}(t)>, \tag{8}$$

where the commutation relations $[\hat{H}_0, \hat{x}] = -i\hbar\mu^{-1}<\hat{p}>$ and $[\hat{H}_0, \hat{p}] = i\hbar\mu\omega_0^2 <\hat{x}>$ have been used [7]. The final vibrational equation of motion associated with the linear Hamiltonian (6) is then derived for the variable $<\hat{x}(t)>$ by substituting (8) into the derivative of equation (7) as

$$\frac{d^2<\hat{x}(t)>}{dt^2} + \omega_{vib}^{(1)2} <\hat{x}(t)> = 0, \tag{9}$$

in which the first-order vibrational frequency $\omega_{vib}^{(1)}$ is the same as the fundamental frequency $\omega_0$, so that we have

$$\omega_{vib}^{(1)} = \omega_0 = \hbar^{-1}\mu c^2 (Z\alpha)^2. \tag{10}$$

Consequently, the complete form of vibrational Hamiltonian HLA (5) is turned out as

$$\hat{H}_{vib} = 0.5\hbar\omega_0 \sum_{k=0}^{\infty} (-1)^{k+1}(k+1)\left(\frac{H_0}{\hbar\omega_0}\right)^k, \qquad (11)$$

in which the expansion coefficients of above series are simultaneously specified by substituting the value of fundamental frequency $\omega_0$ from (10) into (11).

On the other side, the vibrational oscillation frequency of HLA had already been calculated by Bohr as $\omega_n = \frac{m_e}{\mu}\frac{\omega_0}{n^3}$ ($n=1,2,3,...$) in which the electron mass $m_e$ must be substituted by the reduced mass $\mu$ ($m_e = \mu$) as a correction to his classical model because the Hydrogen-like atoms are a two (nuclei +electron) rather a single (electron) particle system [11]. One can probe the complete agreement between our quantum model and Bohr's classical model by expanding the latter oscillation frequency $\omega_n(n) = \frac{\omega_0}{n^3}$ around the lowest value $n$ ($n=1$). The first term of expansion is turned out equal to the fundamental frequency $\omega_0$ given by (10).

## 4. The vibrational Hamiltonian in the nonlinear regime

The second-order nonlinear vibrational Hamiltonian of HLA is assigned by the values $k \leq 2$ in the positive power series (11) as

$$\hat{H}_{vib}^{(2)} = 0.5\hbar\omega_0\left(-1 + \frac{2\hat{H}_0}{\hbar\omega_0} - \frac{3\hat{H}_0^2}{\hbar^2\omega_0^2}\right). \qquad (12)$$

The motion equations of variables $<\hat{x}>$ and $<\hat{p}>$ are similarly rendered by substituting $\hat{H}_{vib}^{(2)}$ for $\hat{H}_{vib}^{(1)}$ in the Heisenberg equations (7) and (8) as

$$\frac{d<\hat{x}(t)>}{dt} = \frac{1}{\mu}\left(1 - \frac{3E_n^0}{\hbar\omega_0}\right)<\hat{p}(t)> - \frac{3}{2}i\omega_0<\hat{x}(t)> \qquad (13)$$

and

$$\frac{d<\hat{p}(t)>}{dt} = -\mu\omega_0^2\left(1 - \frac{3E_n^0}{\hbar\omega_0}\right)<\hat{x}(t)> - \frac{3}{2}i\omega_0<\hat{p}(t)>. \qquad (14)$$

The motion equation for the variable $<\hat{x}>$ is rendered by the respective substitution $<\hat{p}(t)>$ and $d<\hat{p}(t)>/dt$ from (13) and (14) into the derivative of (13) as

$$\frac{d^2<\hat{x}(t)>}{dt^2} + i\beta^{(2)}\frac{d<x(t)>}{dt} + \left(\omega_{vib}^{(2)2} - \frac{\beta^{(2)2}}{4}\right)<\hat{x}(t)> = 0, \qquad (15)$$

in which the second-order vibrational frequency $\omega_{vib}^{(2)}$ is equal to $(4-3n)\omega_0$. Clearly, the second-order term $\beta^{(2)} = 3\omega_0$ must be ignored because it is only a phase term. Therefore, the second-order equation (15) is simplified to the first-order equation (9) with the vibrational frequency equal to $\omega_{vib}^{(2)}$ rather $\omega_{vib}^{(1)}$. In the meantime, the first and second-order expansions of Bohr's vibrational frequency $\omega_n(n) = \dfrac{\omega_0}{n^3}$ around $n = 1$ reveal the complete agreement with the corresponding first and second-order vibrational frequencies $\omega_{vib}^{(1)} = \omega_0$ and $\omega_{vib}^{(2)} = (4-3n)\omega_0$ appeared in the equations (9) and (15), respectively.

As a result, the common features of the first and second-order vibrational Hamiltonians $\hat{H}_{vib}^{(1)}$ and $\hat{H}_{vib}^{(2)}$ defined by (6) and (12) and their corresponding motion equations (19) and (15) [$\beta^{(2)} = 0$] imply that the general Hamiltonian of HLA (11) mimics the motion equation of a quantum harmonic oscillator (QHO) in the general form

$$\frac{d^2 <\hat{x}(t)>}{dt^2} + \omega_{vib}^2 <\hat{x}(t)> = 0, \tag{16}$$

in which

$$<\hat{x}(t)> = <\hat{x}(0)> \left[\cos(\omega_{vib} t) + \sin(\omega_{vib} t)\right] \tag{17}$$

and

$$\omega_{vib} = \frac{\omega_0}{n^3} = \frac{\hbar^{-1}\mu c^2 (Z\alpha)^2}{n^3}. \tag{18}$$

It is reminded that the vibrational level number $n$ takes the values 1, 2, 3, …, and the lowest value $n = 1$ is only used for constructing the vibrational Hamiltonian without appearing in the final results (16)-(18).

## 5. The positional, potential, kinetic, and spontaneous emission noise fluxes

According to the fluctuation-dissipation theorem (FDT), the motion of all oscillators including the vibrational motion of HLA incurs fluctuation when they oscillate in a dissipative medium [13]. Therefore, SPE acts like a damping term in the QHO equation of HLA (16) in the form

$$\frac{d^2 <\hat{x}(t)>}{dt^2} + A \frac{d<\hat{x}(t)>}{dt} + \omega_{vib}^2 <\hat{x}(t)> = 0, \tag{20}$$

in which $A = \tau_R^{-1} = 2\gamma_{sp}$ is Einstein's A-coefficient equal to the inverse of radiative lifetime of the excited state [14]. Equivalently, $2\gamma_{sp}$ is equal to the atomic decay rate of HLA due to the spontaneous emission radiation.

FDT now requires that the fluctuation (Langevin) force $\Gamma_x(t)$ is added to the equation (20) to evaluate the random fluctuation $<\delta\hat{x}(t)>$ that occurs for the position of oscillating electron $<\hat{x}(t)>$ due to the very high vibrational frequency ($\approx 10^{15}\,Hz$) [9, 15]. Finally, the motion equation of fluctuating variable $<\delta\hat{x}(t)>$ is derived as

$$\frac{d^2<\delta\hat{x}(t)>}{dt^2} + 2\gamma_{sp}\frac{d<\delta\hat{x}(t)>}{dt} + \omega_{vib}^2<\delta\hat{x}(t)> = \frac{\Gamma_x(t)}{\mu}, \quad (21)$$

in which $\mu$ is the reduced mass, and $\omega_{vib}$ is the vibrational frequency of electron in HLA as defined in (18). It is reminded that equation (21) is the well-known Langevin equation which here derived for the vibrational motion of HLA for the first time by using the vibrational Hamiltonian of HLA $\hat{H}_{vib}$ given by (11).

The Langevin equation (21) can easily be solved in frequency rather time domain by implementing the following Fourier integral [16]

$$<\delta\hat{x}(\omega)> = \frac{1}{\sqrt{2\pi}}\int_{-\infty}^{+\infty} dt\, <\delta\hat{x}(t)>e^{-i\omega t}, \quad (22)$$

so that the first and second-order temporal differentials $d/dt$ and $d^2/dt^2$ are respectively transformed to $i\omega$ and $-\omega^2$ in frequency domain. As a result, the Fourier transform of (21) gives the following solutions for the fluctuations that imposed to the position $<\delta\hat{x}(\omega)>$ and linear momentum $<\delta\hat{p}(\omega)>$ of oscillating electron

$$<\delta\hat{x}(\omega)> = \frac{\Gamma_x(\omega)/\mu}{-\omega^2 + \omega_{vib}^2 + 2i\gamma_{sp}\omega} \quad (23)$$

and

$$<\delta\hat{p}(\omega)> = \frac{i\omega\Gamma_x(\omega)}{-\omega^2 + \omega_{vib}^2 + 2i\gamma_{sp}\omega}, \quad (24)$$

where the relation $<\delta\hat{p}(\omega)> = i\mu\omega<\delta\hat{x}(\omega)>$ has been used.

Now, the energy conservation relation $<\hat{E}(\omega)>=<\hat{U}(\omega)>+<\hat{K}(\omega)>$ associated with the potential, kinetic and total energies of oscillating electron can be converted to a conservation relation for the corresponding fluctuations as

$$<\delta\hat{E}(\omega)>=<\delta\hat{U}(\omega)>+<\delta\hat{K}(\omega)>. \tag{25}$$

Therefore, our first priority is to calculate the respective potential and kinetic fluctuations $<\delta\hat{U}(\omega)>$ and $<\delta\hat{K}(\omega)>$. By applying the fluctuation $<\delta\hat{x}(t)>$ into the potential energy $<\hat{U}(t)>=0.5\mu\omega_0^2<\hat{x}(t)>^2$, a fluctuation relation is derived for the potential energy correct to first-order approximation ($<\delta\hat{x}(t)>^2=0$) as

$$<\delta\hat{U}(t)>=\mu\omega_0^2<\hat{x}(t)><\delta\hat{x}(t)>. \tag{26}$$

The Fourier transform of (26) led to a relation for its counterpart in frequency domain as

$$<\delta\hat{U}(\omega)>=0.5(1-i)\mu\omega_0^2<\hat{x}(0)><\delta\hat{x}(\omega-\omega_{vib})>, \tag{27}$$

where the solution (17) is substituted into the Fourier transform of (26). Meanwhile, the relation (27) also includes a term proportional to $<\delta\hat{x}(\omega+\omega_{vib})>$ which has been eliminated because it will produce an identical profile with a peak located at the unacceptable negative frequency $\omega=-\omega_{vib}$.

The same procedure should be followed to derive the similar result for the kinetic energy as

$$<\delta\hat{K}(\omega)>=-0.5(1-i)\mu\omega_0<\hat{x}(0)>(\omega-\omega_{vib})<\delta\hat{x}(\omega-\omega_{vib})>, \tag{28}$$

where the fluctuation relation $<\delta\hat{p}(\omega-\omega_{vib})>=i\mu(\omega-\omega_{vib})<\delta\hat{x}(\omega-\omega_{vib})>$ is implemented.

The next important stage is to consider the correlation functions of Langevin force $\Gamma_x(\omega)$ in the forms [6, 17, 18]

$$<\Gamma_x^*(\omega'+\omega_i)\Gamma_x(\omega+\omega_i)>=<\Gamma_x^*(\omega'-\omega_i)\Gamma_x(\omega-\omega_i)>=2\gamma_{sp}\bar{n}\,\delta(\omega-\omega')\approx 0 \tag{29}$$

and

$$\begin{aligned}<\Gamma_x(\omega+\omega_i)\Gamma_x^*(\omega'+\omega_i)>&=<\Gamma_x(\omega-\omega_i)\Gamma_x^*(\omega'-\omega_i)>\\&=2\gamma_{sp}(\bar{n}+1)\delta(\omega-\omega')\approx 2\gamma_{sp}\delta(\omega-\omega')\end{aligned}, \tag{30}$$

in which

$$\bar{n}=\hbar\omega\left[\frac{1}{2}+\left(\exp(\hbar\omega/k_BT)-1\right)^{-1}\right], \tag{31}$$

is the mean number of photons at the frequency mode $\omega$. The first term is associated with the zero-point energy which becomes important at the low temperature conditions $k_B T \ll \hbar\omega$ so that $\bar{n} = \hbar\omega/2$. The second term is due to the thermal radiation and plays significant role at the high temperature conditions $k_B T \gg \hbar\omega$ in order that $\bar{n} = k_B T$. However, the mean number of zero-point and thermal photons is always negligible ($\bar{n} \ll 1$) due to the rather small values $\hbar$ at the low temperature and $k_B$ at the high temperature, as it is evident in the correlation functions (29) and (30).

On the other hand, the correlation function of an arbitrary fluctuating variable $a(\omega)$ with a white noise origin (Dirac function) is defined in the following complex conjugate form [19, 20]

$$<a(\omega)a^*(\omega')> = 2\pi h(\omega) h^*(\omega') \delta(\omega - \omega'), \tag{32}$$

in which $N(\omega) = |h(\omega)|^2$ is the dimensionless mean flux per unit angular frequency bandwidth at angular frequency $\omega$ and given by

$$N(\omega) = \frac{1}{2\pi} \int d\omega' <a(\omega)a^*(\omega')> \exp[i(\omega-\omega')t]. \tag{33}$$

The noise flux spectrum of oscillating electron associated with its positional fluctuations is first calculated by choosing $a(\omega) = <\delta\hat{x}(\omega - \omega_{vib})>$ as

$$N_x(\omega) = \frac{\gamma_{sp}/\pi\mu^2}{\omega^2(\omega - 2\omega_{vib})^2 + 4\gamma_{sp}^2(\omega - \omega_{vib})^2}, \tag{34}$$

where the relations (23), (30), and (33) are used. It is reminded that the fluctuation in the electron position in turn give rises to a fluctuation in its linear momentum according to $<\delta\hat{p}(\omega)> = i\mu\omega <\delta\hat{x}(\omega)>$. Consequently, the fluctuations $<\delta\hat{x}(\omega)>$ and $<\delta\hat{p}(\omega)>$ led to the corresponding fluctuations for the potential and kinetic energies of HLA due to the relations (27) and (28).

The next important task is to calculate the noise fluxes of potential and kinetic energies by substituting $a(\omega) = <\delta\hat{U}(\omega)>$ and $a(\omega) = <\delta\hat{K}(\omega)>$ from (27) and (28) into the noise flux relation (33) as

$$N_U(\omega) = \frac{1}{2}\mu^2\omega_0^4 <\hat{x}(0)>^2 N_x(\omega) \tag{35}$$

and

$$N_K(\omega) = \frac{1}{2}\mu^2\omega_0^2 <\hat{x}(0)>^2 (\omega-\omega_{vib})^2 N_x(\omega). \tag{36}$$

One can now calculate the total noise flux emitted from HLA in the form of spontaneous emission radiation by taking the correlation function of fluctuation conservation relation (25). The noise flux of spontaneous emission $N_{SP}(\omega) = <<\delta\hat{E}(\omega)><\delta\hat{E}(\omega')>^*>$ is thus related to the potential $N_U(\omega) = <<\delta\hat{U}(\omega)><\delta\hat{U}(\omega')>^*>$ and kinetic $N_K(\omega) = <<\delta\hat{K}(\omega)><\delta\hat{K}(\omega')>^*>$ noise fluxes in the following noise flux conservation

$$N_{SP}(\omega) = N_U(\omega) + N_K(\omega) + N_{UK}(\omega) + N_{KU}(\omega), \tag{37}$$

in which $N_{UK}(\omega) = <<\delta\hat{U}(\omega)><\delta\hat{K}(\omega')>^*>$ and $N_{KU}(\omega) = <<\delta\hat{K}(\omega)><\delta\hat{U}(\omega')>^*>$ are the interchange noise fluxes between the potential and kinetic energies. They are turned out to be equal with each other in order that we have

$$N_{UK}(\omega) = N_{KU}(\omega) = -\frac{1}{2}\mu^2\omega_0^3 <\hat{x}(0)>^2 (\omega-\omega_{vib})N_x(\omega). \tag{38}$$

We now consider the atomic transition $2P \to 1S$ of Hydrogen atom as a typical case with the upper energy level $n=2$, the fundamental frequency $\omega_0 = 4.15\times10^{16}$ $Hz$, the vibrational frequency $\omega_{vib} = 5.19\times10^{15}$ $Hz$, and the spontaneous emission (radiative) decay rate $\gamma_{sp} = 4.69\times10^8$ $Hz$ [21]. The noise flux spectrum of electron position $N_x(\omega)$ is plotted in Fig. 1 for the atomic transition $2P \to 1S$. $N_x(\omega)$ is interpreted as the number of fluctuations that occurs for the position of oscillating electron in the unit of time (second) at the oscillation frequency $\omega$. The most physical feature of positional noise flux $N_x(\omega)$ is to play a fundamental role in producing other noise fluxes according to (35)-(38).

The noise flux spectra of potential $N_U(\omega)$ (red), kinetic $N_K(\omega)$ (green), and their equal interchange $N_{UK}(\omega) = N_{KU}(\omega)$ (blue) together with their sum in the form of spontaneous emission radiation $N_{SP}(\omega)$ (black) are illustrated in Fig. 2. The kinetic and interchange noise fluxes are evidently much smaller than the potential noise flux because of their respective proportionality to $(\omega-\omega_{vib})^2$ and $(\omega-\omega_{vib})$ in (36) and (38). Therefore, they have a minimum value around the peak frequency $\omega \approx \omega_{vib}$ from mathematical point of you. The physical interpretation is concerned to the high acceleration of oscillating electron at its oscillation amplitude where the potential energy has the maximum value in contrast with the kinetic energy. The potential energy thus undertakes the maximum uncertainty (noise) due to a sudden change in the oscillation direction of electron. Meanwhile, the negative values for the noise fluxes $N_{UK}(\omega) = N_{KU}(\omega)$ are meaningless except interpreted as the interchanged noise fluxes between the kinetic and potential energies rather submitting to the spontaneous emission radiation.

Finally, the spectrum of spontaneous emission noise flux, which is formed from the linear contribution of four different internal noise fluxes $N_U(\omega)$, $N_K(\omega)$, and $N_{UK}(\omega) = N_{KU}(\omega)$ in Fig. 2, has a Lorentzian profile with the bandwidth equal to $A = 2\gamma_{sp} = 9.38 \times 10^8 \, Hz$ in complete agreement with the experimental work of weak measurement based on atomic spontaneous emission [22].

## 6. Conclusion

A heuristics model is introduced for constructing the vibrational Hamiltonian of a two-particle system in quantum mechanics. This model has already been implemented to form the vibrational [7] and rovibrational [23] Hamiltonians of diatomic molecules. Here, the vibrational Hamiltonian of Hydrogen-like atoms (5) has similarly defined as an infinite positive power series. The series coefficients are simultaneously determined by applying the first-order vibrational Hamiltonian (6) into the Heisenberg equations (7) and (8). The second-order vibrational Hamiltonian (12) clarified the final form of vibrational motion equation of HLA (16) which acts like a quantum harmonic oscillator in the absence of any damping process.

Many physicists have ever tried to explain SPE process by involving the different external sources [1-6]. By contrast, the Langevin equation (21) is here formed to describe the origin of SPE radiation without involving any ambient effect. The solutions of Langevin equation (23) and (24) determined the fluctuations $<\delta \hat{x}(\omega)>$ and $<\delta \hat{p}(\omega)>$ that imposed to the position and linear momentum of oscillating electron in HLA, respectively. These fluctuations are then implemented to calculate the noise fluxes of the internal potential $N_U(\omega)$ (35), kinetic $N_K(\omega)$ (36), and their equal interchange $N_{UK}(\omega) = N_{KU}(\omega)$ (38) by using the correlation functions of Langevin force $\Gamma_x(\omega)$ defined by (29) and (30). It is finally demonstrated that the internal noise fluxes are responsible for the external noise flux of spontaneous emission radiation according to the flux conservation relation (37). The sum of internal noise flux spectra gives a Lorentzian profile for the spontaneous emission radiation with a bandwidth equal to Einstein's A-coefficient, as illustrated in Fig. 2.

In summary, it is demonstrated that the radiation equation of spontaneous emission from atoms obeys from the Langevin equation. The spontaneous emission is formed inside the atom by the fluctuations in the potential, and kinetic energies of oscillating electron in the absence of ambient effects. The next aim is to consider the relativistic motion effect of electron on the spectrum of spontaneous emission radiated from atoms. The calculations are under good progress and the results will be announced in future soon.

**Figure Captions**

**Fig. 1.** The positional noise flux spectrum of oscillating electron $N_x(\omega)$ is demonstrated for the Hydrogen atom in the atomic transition $2P \to 1S$. This noise flux is fundamental because it is responsible for all internal and external noise fluxes due to (35)-(38).

**Fig. 2.** The noise flux spectra of potential $N_U(\omega)$ (red), kinetic $N_K(\omega)$ (green), and their equal interchanges $N_{UK}(\omega) = N_{KU}(\omega)$ (blue) are plotted for the Hydrogen atom in the atomic transition $2P \to 1S$. The sum of these four internal noise fluxes is emerged from the atom as the spontaneous emission radiation with a Lorentzian profile of width $A = 2\gamma_{sp} = 9.38 \times 10^8 \, Hz$ according to the conservation relation (37), as illustrated with the black colour.

Fig. 1.

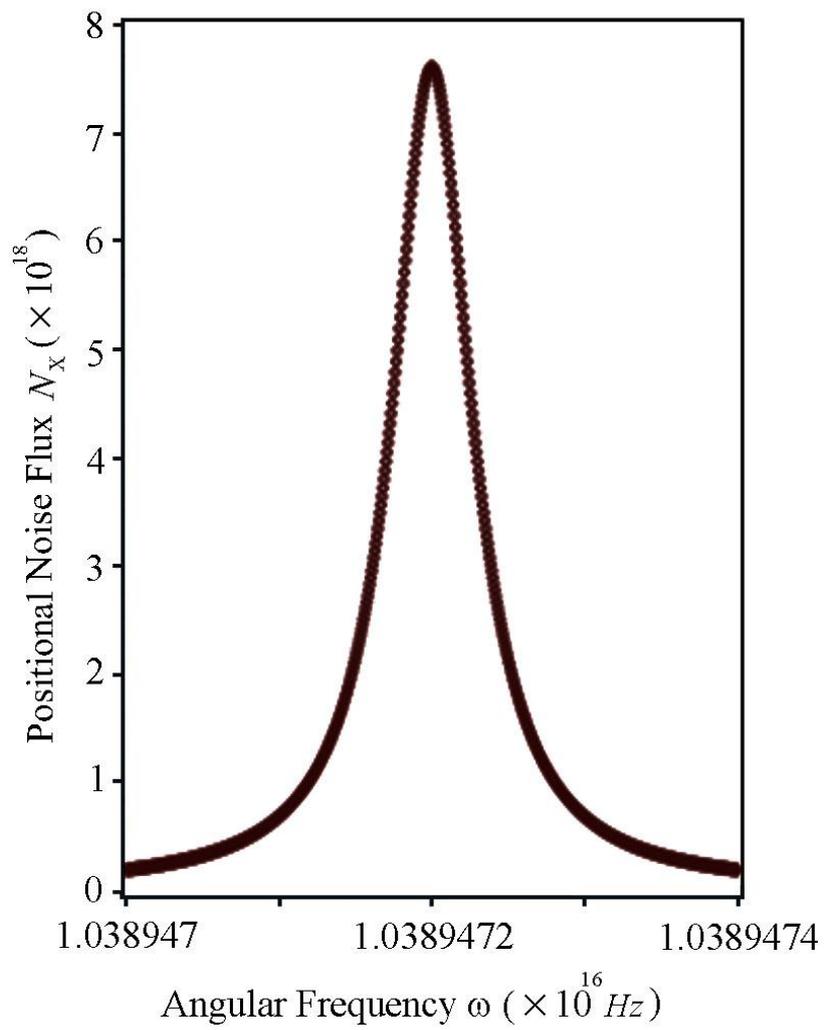

Fig. 2.

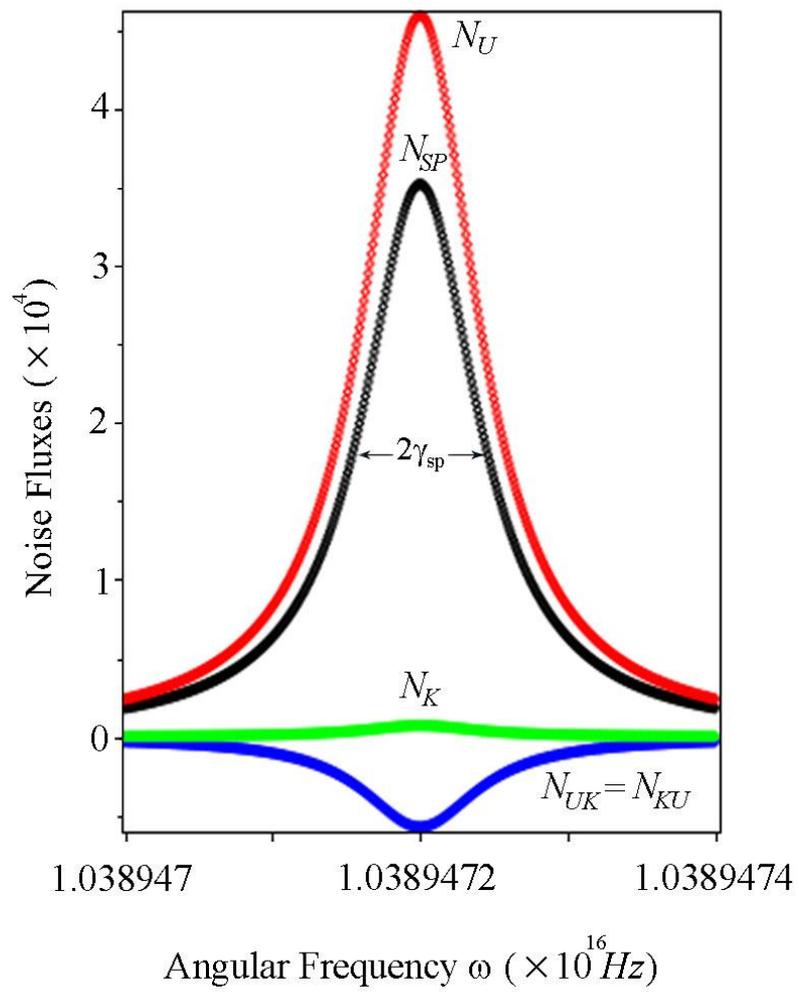